\documentclass[reprint, superscriptaddress, secnumarabic, amssymb, nobibnotes, aps, prl]{revtex4-1}

\usepackage{graphicx}
\usepackage{epstopdf}
\usepackage[T1]{fontenc}
\usepackage[latin9]{inputenc}
\usepackage{amsbsy}
\usepackage{gensymb}
\usepackage[T1]{fontenc}
\usepackage[latin9]{inputenc}
\usepackage{amsmath}
\usepackage{amssymb}
\usepackage{bbm}
\usepackage{physics}
\usepackage{braket}
\usepackage{xcolor}
\allowdisplaybreaks
\usepackage{graphicx}
\usepackage[colorlinks=true]{hyperref}  
\hypersetup{
    bookmarks=true,         
    unicode=false,          
    pdftoolbar=true,        
    pdfmenubar=true,        
    pdffitwindow=false,     
    pdfstartview={FitH},    
    pdftitle={Lu3Os4Ge13},    
    pdfauthor={},     
    pdfsubject={},   
    pdfcreator={},   
    pdfproducer={}, 
    pdfkeywords={} {} {}, 
    pdfnewwindow=true,      
    colorlinks=true,       
    linkcolor=blue, 
    citecolor=blue,        
    filecolor=magenta,      
    urlcolor=blue           
} 
\usepackage[normalem]{ulem}

\newcommand{\equsref}[2]{Eqs.~(\ref{#1}) and (\ref{#2})}

\newcommand{\figref}[1]{Fig.~\ref{#1}}
\newcommand{\refcite}[1]{Ref.~\onlinecite{#1}}

\newcommand{\tableref}[1]{Table~\ref{#1}}

\renewcommand{\approx}{\simeq}

\begin{document}
\title{\textrm{Time-reversal symmetry breaking \\ in superconducting low-carrier-density quasi-skutterudite Lu$_3$Os$_4$Ge$_{13}$}}
\author{A.~Kataria}
\affiliation{Department of Physics, Indian Institute of Science Education and Research Bhopal, Bhopal, 462066, India}
\author{J.~A.~T.~Verezhak}
\affiliation{Department of Physics, University of Warwick, Coventry, CV4 7AL, UK}
\author{O.~Prakash}
\affiliation{Department of Condensed Matter Physics and Materials Science,
Tata Institute of Fundamental Research, Mumbai-400005, India}
\author{R.~K.~Kushwaha}
\affiliation{Department of Physics, Indian Institute of Science Education and Research Bhopal, Bhopal, 462066, India}
\author{A.~Thamizhavel}
\affiliation{Department of Condensed Matter Physics and Materials Science,
Tata Institute of Fundamental Research, Mumbai-400005, India}
\author{S.~Ramakrishnan}
\affiliation{Department of Condensed Matter Physics and Materials Science,
Tata Institute of Fundamental Research, Mumbai-400005, India}
\author{M.~S.~Scheurer}
\affiliation{Institute for Theoretical Physics, University of Innsbruck, A-6020 Innsbruck, Austria}
\author{A.~D.~Hillier}
\affiliation{ISIS Facility, STFC Rutherford Appleton Laboratory, Didcot OX11 0QX, United Kingdom}
\author{R.~P.~Singh}
\email[]{rpsingh@iiserb.ac.in}
\affiliation{Department of Physics, Indian Institute of Science Education and Research Bhopal, Bhopal, 462066, India}
\date{\today}
\begin{abstract}
The complex structure of the Remeika phases, the intriguing quantum states they display, and their low carrier concentrations are a strong motivation to study the nature of their superconducting phases. In this work, the microscopic properties of the superconducting phase of single-crystalline Lu$_3$Os$_4$Ge$_{13}$ are investigated by muon-spin relaxation and rotation ($\mu$SR) measurements. The zero-field $\mu$SR data reveal the presence of spontaneous static or quasi-static magnetic fields in the superconducting state, breaking time-reversal symmetry; the associated internal magnetic field scale is found to be exceptionally large ($\approx$ 0.18~mT). Furthermore, transverse-field $\mu$SR measurements in the vortex state of Lu$_3$Os$_4$Ge$_{13}$ imply a complex gap function with significantly different strengths on different parts of the Fermi surface. While our measurements do not completely determine the order parameter, they strongly indicate that electron-electron interactions are essential to stabilizing pairing in the system, thus, demonstrating its unconventional nature.
\end{abstract}
\maketitle

\section{Introduction}

Intriguing phenomena, such as non-Fermi liquid behaviour, spin or charge order, non-trivial band topology, and a complex crystal structure, often accompany unconventional pairing mechanisms and non-trivial, symmetry-breaking superconducting order parameters in phase diagrams. The Remeika 3-4-13 series is an example of a family of materials, which exhibits a wide array of diverse, exotic phenomena \cite{Rp,cdwrp,mo,qcp,nfl}, and exciting properties, including the Kondo effect, spin-fluctuation superconductivity, multi-gap superconductivity, and many more \cite{sp,ko,musrss}. The Remeika phase has the form R$_{3}$A$_{4}$X$_{13}$, where R is a rare earth metal, A is a transition metal, and X is a group-14 element \cite{Rp1}, and its structure motif is isomorphic to cage-like structures, namely clathrates and filled skutterudites \cite{cla,squ}. The stannides of the 3-4-13 family are believed to be strongly-coupled superconductors with a second-order structural phase transition, pointing towards a strong interplay between the superconducting order and their structure \cite{qcp,qcp1, sepc, musrsp,msrmo,StannidesDisorder}. Meanwhile, R$_{3}$A$_{4}$Ge$_{13}$, which are similar to the stannide structure, are scarce and mostly uninvestigated. A number of germanide compounds exhibit low-temperature 
superconductivity and paramagnetic character different from the stannides \cite{gtc,ypg}; thus presenting an interesting platform in the 3-4-13 family to inspect the connection between crystal complexity and superconductivity.

In the 3-4-13 germanides family, Lu$_{3}$Os$_{4}$Ge$_{13}$, a semimetallic compound, has shown interesting properties in the superconducting state and attracted attention recently. Two initial studies of Lu$_{3}$Os$_{4}$Ge$_{13}$ by Prakash \textit{et al.}~revealed two-gap bulk superconductivity from low-temperature electronic specific heat measurements and a non-linear dependence of the Sommerfeld coefficient, $\gamma_n$, under magnetic field \cite{logtc,logsh}. The superfluid density measurements via tunnel diode oscillator (TDO) also confirmed the two-gap nature of the superconducting state \cite{logtdo}. Another noteworthy property of Lu$_{3}$Os$_{4}$Ge$_{13}$ is the high value of the upper critical field, $H_{c2}$, which is close to its Pauli limiting value \cite{logwhh}, and the high transition temperature $T_c$ for a relatively low carrier density system. Other low carrier density systems with high $T_c$ include cuprates, fullerenes and MgB$_2$, which are known for their fascinating properties \cite{cu,fl,mb2}. 
Exploring the phenomenology of low-carrier-density superconductors is also crucial for our theoretical understanding as the conventional BCS theory is not applicable to the low-density limit.
This motivates a detailed microscopic investigation of the germanide superconductor Lu$_{3}$Os$_{4}$Ge$_{13}$ in order to understand its superconducting order parameter, the pairing mechanism and their relation with the structure in the superconducting phase.
\begin{figure*}[ht!]
\centering 
\includegraphics[width=2.0\columnwidth, origin=b]{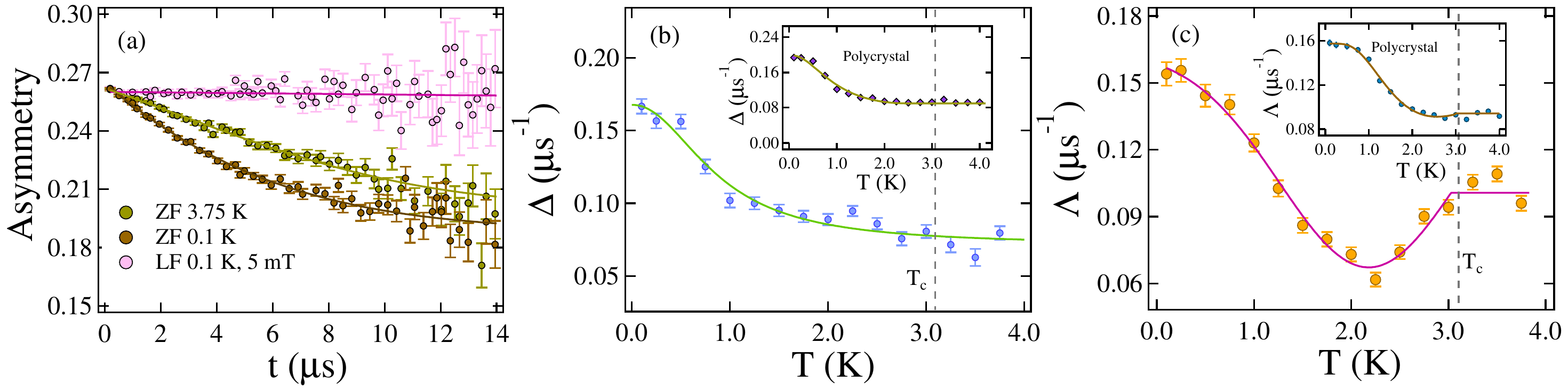}
\caption{\label{Fig1:ZF} (a) ZF muon-spin relaxation spectra for a single crystal of Lu$_3$Os$_4$Ge$_{13}$ at a temperature above (green colour) and below (brown colour) transition temperature with the LF-$\mu$SR spectra recorded in an applied field of 5 mT at 0.1 K. (b) $\Delta$ and (c) $\Lambda$ variation with a temperature estimated by fitting  \equsref{eq1:kt}{eq2:zf} for the single-crystalline sample; the inset of the two represents the same relaxation variation for the polycrystal sample.}
\end{figure*}

Moreover, other cage-type structure systems such as R$_5$Rh$_6$Sn$_{18}$ (R = Sc, Lu and Y) and Pr-based heavy-fermion filled skutterudite (Pr,La)(Ru,Os)$_4$Sb$_{12}$, PrPt$_4$Ge$_{12}$ are among the few compounds which show time-reversal symmetry breaking (TRSB) in their superconducting state with multi-gap features \cite{lrs,srs,yrs,PO4S12,PR4S12,PP4G12}. This further motivates a detailed study of the symmetry-properties of pairing in the quasi-skutterudite Lu$_3$Os$_4$Ge$_{13}$, which exhibits another analogous cage-type structure with multi-gap superconductivity and structural complexity.  

In this paper, the superconducting state of Lu$_3$Os$_4$Ge$_{13}$ is investigated microscopically, using muon spin-rotation and relaxation measurements ($\mu$SR).  Zero-field (ZF) muon spin relaxation measurements reveal a significant variation of the relaxation rate with temperature and, thus, the presence of spontaneous magnetic fields, breaking time-reversal symmetry in the superconducting state. Furthermore, transverse-field (TF) measurements of Lu$_3$Os$_4$Ge$_{13}$ provide the temperature dependence of the superconducting contribution to the relaxation rate, $\sigma_{sc}$, which---in accordance with the previous reports \cite{logsh,logtdo}---is consistent with a multi-gap state with significantly different gap magnitude on different Fermi sheets. We find comparatively small values of $\Delta_i (T=0)/k_B T_c$ for the two superconducting gaps, $\Delta_i$, indicative of weakly coupled superconductivity in Lu$_{3}$Os$_{4}$Ge$_{13}$. 

\section{Experimental Details}

Both single and poly-crystalline samples of Lu$_{3}$Os$_{4}$Ge$_{13}$ have been used for $\mu$SR measurement; these samples have already been characterized and previously studied and reported in refs. \cite{logtc,logsh}. The phase purity of the sample is investigated by powder x-ray and Laue-diffraction pattern. Resistivity and magnetization confirmed the superconducting onset at temperature $T_c$ = 3.1 K. $\mu$SR measurements were performed in various configurations, including zero-field, longitudinal-field (LF) and transverse-field mode using the $\mu$SR spectrometer at the ISIS Neutron and Muon Pulsed Source, Appleton Laboratory, United Kingdom. Three sets of orthogonal coils and an active compensation system are present in the spectrometer to cancel or remove any stray magnetic field at the sample position.  A detailed description of the technique can be found in the refs.\cite{isis1, isis}. TF measurements were performed in the superconducting mixed or vortex state with an applied magnetic field perpendicular to the muon spin direction to evaluate the superconducting gap structure. Moreover, the ZF and LF measurements in the longitudinal geometry will probe the presence of weak spontaneous magnetic fields and local magnetism in the respective state.

\section{Results}

\subsection {Zero-Field $\mu$SR}

\begin{figure*}
\centering
\includegraphics[width=2.0\columnwidth, origin=b]{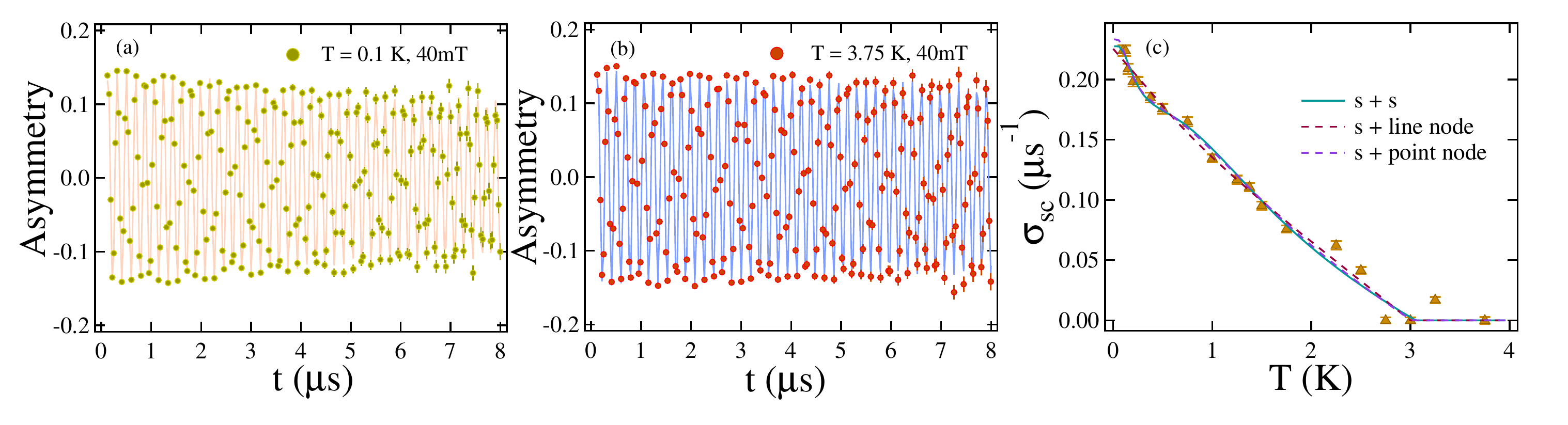}
\caption{\label{Fig2:TF} The time-domain TF-$\mu$SR spectra of single-crystalline sample in the presence of 40mT magnetic field (a) below (superconducting state) and (b) above (normal state) the transition temperature. (c) The muon spin depolarization rate, $\sigma_{sc}$ variation with the temperature where the cyan solid line, pink dotted line and purple dotted line represent the (a) $s + s$ (b) $s + \text{line}$ and  (c) $s + \text{point}$ node, respectively.} 
\end{figure*}

The ZF asymmetry spectra above and below the superconducting transition temperature, together with the LF spectra at 0.1 K and 5 mT magnetic field for single-crystalline Lu$_{3}$Os$_{4}$Ge$_{13}$ are shown in \figref{Fig1:ZF} (a). A significant change in relaxation rate from superconducting (0.1 K) to the normal state (3.75 K) is observed with no oscillating signal, which directs toward the possible presence of a spontaneous magnetic field below the superconducting transition temperature. For static, randomly oriented nuclear moments, the asymmetry spectra can be understood from the Gaussian Kubo-Toyabe (KT) equation \cite{kt},
\begin {equation}
G_z(t)= \frac{1}{3} + \frac{2}{3}(1-\Delta^2t^2)\exp\left(-\frac{\Delta^2t^2}{2}\right)
\label{eq1:kt}
\end {equation}
where $\Delta$ corresponds to relaxation due to the nuclear dipolar field.  The measured time-dependent ZF asymmetry spectra are well described by the muon relaxation function \cite{zff},
\begin{equation}
     A(t)= A_0 G_z(t)\exp(-\Lambda t)+ A_{bg}
     \label{eq2:zf}
\end{equation}
where $A_0$ is the initial asymmetry corresponding to the sample, $A_{bg}$ considers the background asymmetry and $\Lambda$ accounts for the electronic relaxation rate. The temperature dependence of the relaxation rates (i.e., the fitting parameters) $\Delta$ and $\Lambda$ for a single crystal of Lu$_{3}$Os$_{4}$Ge$_{13}$ are shown in \figref{Fig1:ZF} (b) and (c), where the sample and background asymmetries are temperature independent. A significant variation in both $\Lambda$ and $\Delta$ is observed below the transition temperature. The increment in $\Delta$ value is pronounced in the low-temperature region much below $T_c$. The insets of \figref{Fig1:ZF} (b) and (c) depict the similar variation of relaxation rates for the polycrystalline sample.

The increment nature of relaxation rates, $\Lambda$ and $\Delta$, with temperature (\figref{Fig1:ZF} (b) and (c)), can be attributed to either fast fluctuations arising from electronic spins or the presence of a static or quasi-static magnetic field in the sample below the superconducting transition temperature. Considering the case of relaxation due to fluctuations, the field fluctuations perpendicular to an applied field can cause spin-flip transitions of the muon spin, which leads to an overall relaxation of the muon ensemble and a Curie-Weiss-like temperature dependence \cite{sfref}. Further, the slow time evolution of LF asymmetry spectra (\figref{Fig1:ZF} (a)), i.e., the low relaxation rate, indicates the decoupling of muon spin from the local magnetic field environment even with an applied magnetic field as low as 5~mT. This implies that the observed signal of increased relaxation rate with temperature is emerging from a dilute static or quasi-static magnetic field in the system, breaking time-reversal symmetry and excluding the possibility of any extrinsic magnetic impurity effects in Lu$_3$Os$_4$Ge$_{13}$. Such a change in $\Lambda$ has only been observed in a limited set of superconductors, including Sr$_2$RuO$_4$ \cite{S2RO4}, LaNiC$_2$ \cite{trs}, SrPtAs \cite{SPA}, (Lu/Y/Sc)$_5$Rh$_6$Sn$_{18}$ \cite{lrs,yrs,srs}, Ba$_{1-x}$K$_x$Fe$_2$As$_2$ \cite{BKFA}, La$_7$X$_3$ (X = Rh,Ir,Pd,Ni) \cite{L7I3,L7N3,L7P3,L7R3} series and recently in monochalcogenide ScS \cite{ScS}. Furthermore, the $\Delta$ relaxation channel, representing the correlated nuclear dipolar moments of the atoms, has a variation in the low-temperature region significantly below the superconducting transition temperature. The long-baseline value of $\Delta$ is most likely due to Lu atoms having the largest nuclear moment among the three constituent atoms. Besides, the trend in the secondary channel, $\Delta$ here, is also observed in La$_7$Ir$_3$, La$_7$Pd$_3$, La$_7$Rh$_3$ \cite{L7I3,L7P3,L7R3} and Pr$_{1-x}$La$_{x}$Pt$_4$Ge$_{12}$, Pr(Os$_{1-x}$Ru$_x$)$_4$Sb$_{12}$, Pr$_{1-y}$La$_y$Os$_4$Sb$_{12}$ \cite{PLP4G12,POR4S12}. The proposed reason for the increment in $\Delta$ is nuclear spin fluctuations, though the net change observed in $\Delta$ for the aforementioned materials is much lower than the value detected for our single-crystalline Lu$_3$Os$_4$Ge$_{13}$.

The value of static or quasi-static magnetic field present in the sample below the transition is evaluated from the observed change in $\Lambda$. Subtracting the minimum value of the relaxation rate to its maximum value at the lowest temperature yields the net increase $\delta\Lambda$ = 0.151 $\mu$s$^{-1}$. The characteristic local magnetic field strength can be estimated from $\delta\Lambda/\gamma_{\mu} = B_{loc}/\sqrt{2}$, providing $B_{loc} \approx$ 0.18 mT \cite{lmf}. The observed local field strength is much larger than the values reported for the other TRSB compounds in their superconducting state (e.g., for Sr$_2$RuO$_4$ it is 0.005 mT \cite{S2RO4}) but comparable to the value reported for filled skutterudite and frustrated superconductor (0.12 mT for PrOs$_4$Sb$_{12}$ and 0.116 mT for Re$_2$Hf \cite{PO4S12,re2hf}).

Notably, the increment in electronic relaxation rate, $\Lambda$, is observed at a temperature different from the superconducting transition temperature, defined as the onset temperature of time-reversal symmetry breaking, $T_{\text{onset}}$ = 2.2 K. With increasing temperature, a small dip-like feature in the $\Lambda$ is followed by the constant value above the transition temperature. The similar dip-like feature in relaxation channel with two temperatures, $T_{\text{onset}}$ and $T_c$, are also observed in the other TRSB skutterudites superconductors such as PrPt$_4$Ge$_{12}$ \cite{PP4G12}, Pr$_{1-x}$Ce$_{x}$Pt$_4$Ge$_{12}$ \cite{PCP4G12}, Pr$_{1-x}$La$_{x}$Pt$_4$Ge$_{12}$ \cite{PLP4G12} and recently in La$_7$Ni$_3$ \cite{L7N3}. However, the exact reason behind the dip feature is not known and is believed to be associated with the multi-component nature of the superconducting order parameter. For our case of quasi-skutterudite Lu$_3$Os$_4$Ge$_{13}$, this is discussed later. 

\subsection {Transverse-Field $\mu$SR}

\begin{table*}[ht]
\caption{Summary of two superconducting gap analyses for single-crystalline Lu$_3$Os$_4$Ge$_{13}$ via various models from different studies (indicated in the last column), including our TF-$\mu$SR, specific heat (SH), and TDO measurements. }
\begingroup
\begin{ruledtabular}\label{TableWithFits}
\begin{tabular}[b]{c c c c c c c c c}
Model & $\chi^2$ & $T_c$ & fraction $(w_1)$ & $\Delta_{0,1}$ (meV) & $\Delta_{0,2}$ (meV) & $ \Delta_{0,1}/\Delta_{0,2}$ & Measurement & Reference \\ \hline
$s + s$ & 1.36 &3.0(1) &  0.27(4)& 0.04(1)& 0.30(2)& 7.5 & TF & This work\\
$s + \text{point}$ & 1.37 &3.0(1) & 0.24(7) &0.03(1)& 0.37(3)& 12.3 & TF  & This work\\
$s + \text{line}$ & 1.38 &3.0(1) & 0.3(2) &0.6(1)& 1.4(4) & 4.6 & TF  & This work\\
$s + s$ & 1.31 & 3.1 & 0.18& 0.04(3) & 0.43(1) & 10.8 & SH  & \cite{logsh} \\
$s + s$  & & 3.0 & 0.22 & 0.33 & 0.65 & 1.9 & TDO   & \cite{logtdo}
\end{tabular}
\end{ruledtabular}
\par\medskip\footnotesize
\endgroup
\end{table*}

The asymmetry spectra in TF configuration above and below the transition temperature, $T_c$ under an applied magnetic field of 40 mT for single-crystalline Lu$_{3}$Os$_{4}$Ge$_{13}$ are shown in \figref{Fig2:TF} (a) and (b). The decay in asymmetry spectra amplitude below $T_c$ reveals the presence of the inhomogeneous fields of the flux-line lattice. The superconducting gap structure of Lu$_{3}$Os$_{4}$Ge$_{13}$ has been investigated by temperature variation of $\sigma_{sc}$, estimated from the second-moment method \cite{sm}. To this end, the time-domain asymmetry spectra are described by a Gaussian-damped oscillatory muon spin relaxation function expressed as \cite{tff,tff1},
\begin{equation}
A (t) = \sum_{i=1}^N A_{i}\exp\left(-\frac{1}{2}\sigma_i^2t^2\right)\cos(\gamma_\mu B_it+\phi)
\label{eqn3:TF}
\end{equation}
where $A_i$ and $\sigma_i$ are the corresponding initial asymmetry and Gaussian relaxation rate. $\phi$ is the offset phase, and $B_i$ is the $i^{th}$ component of the magnetic field distribution with $\gamma_{\mu}$/2$\pi$ = 135.5 MHz/T being the muon gyromagnetic ratio. Here, $N$ = 2 describes the distribution appropriately, with $\sigma_2$ fixed to zero to account for the non-depolarizing background originating from the sample holder. Thus, $A_2$ exhibits the background asymmetry character, and $B_2$ corresponds to the background magnetic field.

The extracted temperature-dependent total Gaussian relaxation rate, $\sigma$ consists of a contribution from both the flux-line lattice ($\sigma_{sc}$) and nuclear moment ($\sigma_n$). A temperature-invariant relaxation rate from the nuclear moment, $\sigma_n$, is obtained from the asymmetry spectra measured above $T_c$. Thus, the superconducting contribution of the relaxation rate is evaluated via $\sigma_{\mathrm{sc}} = \sqrt{\sigma^{2} - \sigma_{n}^{2}}$. For a vortex lattice system having $\kappa \geq$ 5, and in an applied field much smaller than the upper critical field ($H_{\text{app}} \ll H_c$), the relation of $\sigma_{sc}$ and penetration depth, $\lambda$ reads as \cite{tfs},
\begin {equation}
\frac{\sigma_{sc} (T)}{\gamma_{\mu}} = 0.0609 \frac{\Phi_0}{\lambda^2(T)} 
\label{eqn4:si}
\end {equation}
where $\Phi_0$ is the magnetic flux quantum. Hence, $\sigma_{sc}$ contains the information of the superconducting gap structure. The increase of $\sigma_{sc}$ with decreasing temperature is ascribed to the development of the vortex lattice as the superconductor enters the mixed phase. In \figref{Fig2:TF}(c), the $\sigma_{sc}$ versus $T$ curve is shown where the superconducting transition appears to be very broad. The curve depicts an approximately linear behaviour down to very low temperature and without any saturation features; this indicates nodal pairing or multiple superconducting gaps with significantly different magnitudes while ruling out a single isotropic gap. To be more quantitative, the $\sigma_{sc}$ temperature evolution can be presented in the semiclassical approximation as \cite{lr1,lr2}, 
\begin {multline}
\delta\sigma_{sc}(T,\Delta_{0,i}) = 1 + \\
2\left\langle \int_{\Delta(T,\phi,\theta)}^\infty \frac{\delta f}{\delta E}\frac {E dE}{\sqrt{E^2-|\Delta_i(T,\phi,\theta)|^2}}\right\rangle_{\text{FS}},
\label{eq5:lam}
\end{multline}
here $f(E)= [\exp(E/k_B T)+1]^{-1}$ is the Fermi-Dirac function and $\langle... \rangle_{\text{FS}}$ represents the average over the Fermi surface. The gap function $\Delta_i(T,\phi,\theta)$ consists of the product $\Delta_{i}(T/T_c)g(\phi,\theta)$, with $g(\phi,\theta)$ encoding the angular dependence of the gap around the Fermi surface at an azimuthal angle $\phi$ and polar angle $\theta$. 
We use \cite{2g1}
\begin{equation}
\Delta_{i}(T/T_c) = \Delta_{0,i}\tanh \left\{1.82\left[1.018\left(\frac{T_c}{T}-1\right)\right]^{0.51}\right\}
\label{eqn6:del}
\end{equation}
with $\Delta_{0,i}$ being the gap value at $T = 0$, which approximates the BCS temperature dependence of $\Delta_i/\Delta_{0,i}$ well. To account for the experimental data quantitatively, various two-gap scenario based on the $\alpha$-model \cite{2g2,a-m} are applied: (a) a fully gapped $s + s$ state, (b) a model with a full gap on one and with nodal lines on the second Fermi sheet (denoted by $s+\text{line}$), and (c) a full gap on one and point nodes on the second sheet ($s+\text{point}$). For the full gap, line nodes, and point nodes we use $g(\phi,\theta)$ = 1, $g(\phi,\theta)$ = $|\cos(2\phi)|$, and $g(\phi,\theta)$ = $|\sin \theta|$, respectively.
The two-gap model is incorporated by using a weighted sum of two gap values as \cite{L7N3,lf3s3,lgo},
\begin {equation}
\frac{\sigma_{sc}(T)}{\sigma_{sc}(0)} = w_1  \delta\sigma_{sc}(T,\Delta_{0,1}) + w_2  \delta\sigma_{sc}(T,\Delta_{0,2})\\   
\label{eq7:2g}
\end{equation}
where $w_1$ and $w_2$ are the weighted fractions, respectively, of their superconducting gap $\Delta_{0,1}$ and $\Delta_{0,2}$, with $w_1 + w_2$ = 1. The two-gap model describes the data well but requires significantly different gap magnitudes $\Delta_{0,1}$ and $\Delta_{0,2}$. For instance, the $s + s$ model yields $\Delta_{0,1}(0)$ = 0.04(1) meV and $\Delta_{0,2}(0)$ = 0.30(2) meV with $w_1$ = 0.27(4). The obtained value from the $s + s$-wave model with the lowest $\chi^2$ value agrees with the two-gap values stated from the specific heat data but deviates from the superfluid density measurement values \cite{logtdo}. The superconducting gap analysis of Lu$_{3}$Os$_{4}$Ge$_{13}$ is summarised in Table \ref{TableWithFits} with the observed values of other studies. The variation in gap values via different techniques might be due to the incorporation of various assumptions in the calculations. 
Importantly, since the $\chi^2$ values for the other two-gap models we investigated, $s+\text{line}$ and $s+\text{point}$ nodes, are close, our data does not allow us to distinguish between nodal multi-band pairing and fully gapped multi-gap behavior with strongly varying gap magnitude (near nodal). We note that multi-gap behavior is natural for a system such as Lu$_3$Os$_4$Ge$_{13}$ with several complex Fermi surfaces as revealed by first-principle band structure calculations \cite{logsh} and is also evident from various experimental observations \cite{logsh,logtdo}.
The low dimensionless gap values $\Delta_{0,i}/k_B T_c$ obtained for Lu$_3$Os$_4$Ge$_{13}$ $(\approx 1.1, 0.15)$ are counter to those of the stannides where values above that of the BCS theory have been observed \cite{stannidegapvalue}. 

\section{Discussion}
Lu$_3$Os$_4$Ge$_{13}$ crystallizes in the space group $Pm\bar{3}n$ (no.~223), with point group $O_h$. If the superconducting phase in the system is reached by a single, continuous phase transition, we know that the pairing state must transform under one of the irreducible representations (IRs) of $O_h$ (setting aside rather exotic scenarios where its order parameter transforms non-trivially under lattice translations). In total, $O_h$ has 10 IRs, some of which are two or even three-dimensional, leading to 26 possible superconducting instabilities \cite{tsc}. Our ZF-$\mu$SR measurements, however, indicate TRSB superconductivity, which strongly constrains the order parameter symmetries: first, time-reversal symmetry can only be broken at a single superconducting transition if the order parameter transforms under a multi-dimensional IR. In our case, this leaves us with the six IRs $E_\mu$, $T_{1\mu}$, and $T_{2\mu}$, $\mu=g,u$. Furthermore, only some states transforming under these IRs will break time-reversal symmetry; a systematic analysis shows \cite{tsc} that there are $10$ such TRSB superconductors. Interestingly, symmetry implies that all of these states will either have nodal points or lines on some of the Fermi sheets \cite{logsh}. As already pointed out above [cf.~\figref{Fig2:TF}(c) and \tableref{TableWithFits}], our TF-$\mu$SR results are not incompatible with the nodes in the superconducting states. 

Recall, though, that a fully gapped $s+s$ state is also consistent with our TF-$\mu$SR data, albeit with significantly different gap magnitude on different Fermi surfaces in order to reproduce the seemingly non-exponential behavior of $\sigma_{\text{sc}}$ at low $T$ in \figref{Fig2:TF}(c). In combination with the fact that previous TDO measurements \cite{logtdo} found exponential low-$T$ behavior of the penetration depth of Lu$_3$Os$_4$Ge$_{13}$, we here also discuss how full-gap superconductivity could be reconciled with broken time-reversal symmetry. The above discussion of pairing states makes the simplifying assumption that the TRSB superconductor is reached via a single transition. While we currently do not have any clear indications of several transitions close to $T_c$, the fact that $\Lambda$ in \figref{Fig1:ZF}(c) only increases at a temperature that is significantly smaller than $T_c$ might point towards the following scenario: right at $T_c$, as identified in transport, a first superconducting order parameter sets in; it does not break time-reversal symmetry and may or may not be fully gapped. At a temperature lower than $T_c$, a secondary order parameter sets in and time-reversal symmetry is broken, e.g., due to some non-trivial complex phase between the primary and secondary order parameter which leads to spontaneous fields \cite{SpontFieldssis}. In this way, one can obtain a fully-gapped, TRSB superconductor. The simplest such scenario is an $s+ is$ state, where both superconducting components are trivial under $O_h$ (IR $A_{1g}$) and, hence, generically fully-gapped. The relative complex phase can arise from ``frustrated'' Cooper-channel interactions between the different pockets of the system; see, e.g., \refcite{L7N3} for an illustration in a toy model or \refcite{sispnictides}. Another, more complex, scenario is that the primary and secondary order parameters are even and odd in frequency. As recently pointed out in \refcite{EvenOddFreqMixing}, this can be realized for strong electron-phonon coupling, competing with Coulomb repulsion, if the phonon and Fermi energy scales become comparable. Due to the low carrier concentrations in superconducting Lu$_3$Os$_4$Ge$_{13}$ \cite{logwhh}, this scenario does not seem implausible either.

We emphasize that all of the above scenarios require a repulsive Coulomb interaction on top of electron-phonon coupling to stabilize the pairing phase. As such, our data clearly indicates the presence of an unconventional pairing mechanism in Lu$_3$Os$_4$Ge$_{13}$. Together with the multigap and multiband nature as well as the low carrier concentration, this makes Lu$_3$Os$_4$Ge$_{13}$ an exciting system to study pairing beyond the BCS paradigm. 
 
\section{Summary and Conclusion}
In conclusion, the microscopic properties of the superconducting state are investigated for the low carrier, quasi-skutterudite Lu$_3$Os$_4$Ge$_{13}$. The ZF-$\mu$SR data indicate spontaneous time-reversal-symmetry breaking in the superconducting state, reflected by the increase in relaxation rate $\Lambda$ in \figref{Fig1:ZF}(c)  at a temperature, $T_{\text{onset}} \approx$ 2.2K, below the superconducting transition temperature $T_c \approx 3.1\,\textrm{K}$. The measured strength of the dilute magnetic field in the superconducting state of Lu$_3$Os$_4$Ge$_{13}$ is $\approx$ 0.18mT---a value much larger than that of most TRSB superconductors, while only those of PrOs$_4$Sb$_{12}$ and Re$_2$Hf are comparable yet still smaller \cite{PO4S12,re2hf}. 
The variation of $\sigma_{sc}$ with temperature, extracted from our TF measurements, is inconsistent with a single isotropic pairing state. Meanwhile, it is well described by a two-gap scenario, with large gap anisotropy $\Delta_{0,1}/\Delta_{0,2} \approx 5-12$, in accordance with previous studies \cite{logsh,logtdo} and with the presence of multiple bands at the Fermi level \cite{logsh}; we emphasize that our data is consistent with both fully gapped and nodal states, see \tableref{TableWithFits} and \figref{Fig2:TF}(c). 
A symmetry analysis of pairing states shows that reaching a TRSB superconducting state in Lu$_3$Os$_4$Ge$_{13}$ via a single continuous phase transition is only consistent with the latter, nodal-pairing scenario. However, a full gapped, TRSB state could be naturally reached via two consecutive transitions.

Taken together our findings establish Lu$_3$Os$_4$Ge$_{13}$ as an exciting superconducting compound that combines not only pairing at low carrier concentrations and multiple bands but also spontaneous symmetry-breaking in the superconducting state and unconventional pairing mechanisms.  
More theoretical and experimental work is required on Lu$_3$Os$_4$Ge$_{13}$, in particular, and Remeika phase germanides, in general, to identify the complex microscopic physics in these systems.

\section*{acknowledgements}
R.~P.~S.\ acknowledge Science and Engineering Research Board, Government of India, for the Core Research Grant CRG/2019/001028. We thank ISIS, STFC, UK for the beamtime to conduct the $\mu$SR experiments. M.S.S.~acknowledges funding by the European Union (ERC-2021-STG, Project 101040651---SuperCorr). Views and opinions expressed are however those of the authors only and do not necessarily reflect those of the European Union or the European Research Council Executive Agency. Neither the European Union nor the granting authority can be held responsible for them.

\end{document}